\documentclass[12pt,aps,prd,superscriptaddress,nofootinbib, preprintnumbers]{revtex4-1}

\usepackage[english]{babel}
\usepackage[latin1]{inputenc}
\usepackage{hyperref}
\usepackage{graphicx,color}
\usepackage{latexsym}
\usepackage{amsmath}
\usepackage{amsfonts,dsfont}
\usepackage{amssymb,slashed}
\usepackage{tensor}
\usepackage{bbold}
\usepackage{verbatim,epigraph}

\begin{document}

\title{Localization in the Rindler Wedge}

\author{\textbf{M. Asorey}\footnote{asorey@unizar.es} }

\affiliation{Departamento de F\'isica T\'eorica, Facultad de Ciencias, Universidad de 
Zaragoza, 50009 Zaragoza, Spain}

\author{\textbf{A. P. Balachandran}\footnote{balachandran@gmail.com} }

\affiliation{Department of Physics, Syracuse University, Syracuse, N. Y. 13244-1130, USA}

\author{\textbf{G. Marmo}\footnote{marmo@na.infn.it} }

\affiliation{Dipartimento di Scienze Fisiche, Universit\`a di Napoli Federico II and INFN 
Sezione di Napoli, Via Cintia, 80126, Napoli, Italy}

\author{\textbf{A. R. de Queiroz}\footnote{amilcarq@unb.br}}  

\affiliation{Instituto de Fisica, Universidade de
Brasilia, C. P. 04455, 70919-970, Brasilia, DF, Brazil}




\begin{abstract}
One of the striking features of QED  is that charged particles create a coherent cloud of 
photons. The resultant coherent state vectors of photons generate a 
non-trivial representation of the localized algebra of observables that do not support 
a  representation of the Lorentz group: Lorentz symmetry is spontaneously broken. We show 
in particular that Lorentz boost generators diverge in this representation, a result 
shown also in \cite{Balachandran2015} (See also \cite{Balachandran2013e}). Localization 
of observables, for example in the Rindler wedge, uses Poincar\'e invariance in an 
essential way \cite{Bal2016}. Hence in the presence of charged fields, the photon 
observables cannot be localized in the Rindler wedge. 

These observations may have a bearing on the black hole information loss paradox, as the 
physics in the exterior of the black hole has points of resemblance to that in the Rindler 
wedge.
\end{abstract}

\maketitle

\section{Introduction}

In Mikowski space, the vacuum state is known to become thermal or KMS for massive neutral 
fields restricted to a Rindler wedge. These fields are associated with uniformly 
accelerated particles. If the acceleration is in the $1$-direction, the thermal or 
modular Hamiltonian is the boost $K_1$ in the $1$-direction. We argue that if the fields 
are charged, $K_1$ diverges and in fact all components $K_i$ of ${\bf K}$ diverge. The 
reason is that the photon vacuum becomes dressed with an 
infrared cloud and breaks Lorentz invariance. Charged observables cannot thus be 
localized in the wedge.

The work of \cite{Balachandran2015} also shows a similar divergence of boosts (See also 
\cite{Balachandran2013e}). But the emphasis in that paper is on the breakdown of Lorentz 
invariance and not on localisation problems as in this paper. Also the in state vector 
considered here is different from the state vector considered there for showing this 
divergence.

A consequence of this result is that the standard Tomita-Takesaki theory for the 
``symplectic'' localization of observables \cite{Bal2016} in a Lorentz covariant manner 
breaks down for charged fields.

These results may have a bearing on the information loss paradox for black holes.

Elsewhere \cite{ABLM} we have argued that equations of motion of 
electromagnetic fields generated by charged particles cannot be 
localized in the Rindler wedge because the charged particle itself is not localized.

\section{The Rindler Wedge for Neutral Fields}

The standard Rindler wedge $W_1$ in Minkowski space $M_4$ is the submanifold 
\begin{equation}
  \label{Rindler-wedge-1}
 W_1=\{x=(x^0, x^1, x^2, x^3)\in M_4: ~x^1\geq |x^0| \}
\end{equation}
Its causal complement is the opposite wedge $W_1'$ (prime denoting causal complement),
\begin{equation}
  \label{Rindler-wedge-complement-1}
 W_1'=\{x\in M_4:~-x^1\geq |x^0|\}. 
\end{equation}
For neutral free fields, there is a rigorous theory of localization in such wedges 
(and their intersections. See \cite{Bal2016} and references therein.). It associates 
algebras of 
local observables $A_W$ and $A_{W'}$ of $W$ and $W'$, respectively, compatibly with 
Poincar\'e covariance and causality. Thus this theory incorporates covariance and 
causality.

This theory of localization, called ``modular localization'', is based in particular on 
the representation of the Poincar\'e group on the quantum fields. The construction of 
$A_{W_1}$ for example uses the boost generator $K_1$.

If there are charged fields and their photons, then because of infrared effects, Lorentz 
group is spontaneously broken \cite{Frolich1979}. In particular, we shall see that $K_1$ 
diverges. The 
implications is that localizations in $W$ and $W'$ break down.

From another point of view \cite{ABLM}, we have argued that equations of motion of 
charged 
field cannot be localized in $W$. We suspect that these results have implications for the 
black hole information paradox.

\section{On Modular Localization}

In non-relativistic quantum physics, given the spatial regions $O_1$ and $O_2$ at a fixed 
time with $O_1\cap O_2=\varnothing$, we have projection operators $P_1$ and $P_2$ such 
that $P_1P_2=0$. Hence it is enough to set $\psi_1=P_1 \chi,~\psi_2=P_2 \chi'$ for 
generic wave functions $\chi,\chi'$ to see that there are wave functions $\psi_1$ and 
$\psi_2$ localized in $O_1$ and $O_2$ which are orthogonal, $\langle 
\psi_2|\psi_1\rangle=0$. Such a localization is known as ``Born localization''.

Let us next turn to relativistic quantum field theory and assume for the rest of this 
section that there 
are no infrared effects. Let us also denote by $W$ the standard Rindler wedge 
(\ref{Rindler-wedge-1}), and by $W'$ 
its causal complement (\ref{Rindler-wedge-complement-1}).

As discussed by many authors \cite{Bal2016}, in relativistic physics, we cannot localize 
states. 
We can only localize algebras of observables in the ``symplectic'' or ``modular'' sense. 
That means the following in the present context: we can associate algebras of observables 
$A_W$ and $A_{W'}$ to $W$ and $W'$ which are compatible with causality, that is, if 
$\psi_W$ and $\psi_{W'}$ are elements 
of $A_W$ and $A_{W'}$, then $[\psi_W,\psi_{W'}]=0$. This association is also compatible 
with covariance as we presently discuss.

Thus in modular localization theory, we have a family of spacetime regions $O_i$ to which 
one assigns the algebras of observables $A_{O_i}$. The regions $O_i$ are obtained from 
$W$ 
and $W'$ by transforming them by the elements of the Poincar\'e group 
$\mathcal{P}_+=\{g\}$ 
consisting of the
connected Poincar\'e group and CPT and then by taking all their intersections. The 
algebras of 
observables $A_{O_i}$ are such that we have 
\begin{enumerate}
 \item \emph{covariance}: we have a representation $g\to U(g)$ of the Poincar\'e group 
$\mathcal{P}_+$ such that if $g\cdot O$ is the Poincar\'e transform of $O$, then 
$A_{g\cdot O}=U(g) A_O U(g)^{-1}$;
\item \emph{causality}: the algebra $A_{O'}$ is the commutant $A_O'$ of $A_O$;
\item \emph{isotony}: if $O_1\subset O_2$, then $A_{O_1}\subseteq A_{O_2}$ (We will not 
discuss isotony further)
\end{enumerate}

For our purposes in this paper, it is enough to consider $A_W$ and $A_{W'}$. Let us first 
consider $A_W$ and a free massive real scalar field $\varphi$. Let $\{f_W\}$ be a 
collection of smooth real test functions supported on $W$. Then the transformation
\begin{equation}
\label{JW-operator}
 \mathcal{J}_W:(x^0,x^1,x^2,x^3)~\to ~ (-x^0,-x^1,x^2,x^3)
\end{equation}
transforms $\{f_W\}$ to the test functions $\{f_{W'}\}=\{\mathcal{J}_W f_W\}$ supported 
in $W'$. In quantum theory $\mathcal{J}_W$ becomes
\begin{equation}
 U(\mathcal{J}_W)\equiv J_W = ~\text{CPT}\times \pi\text{-rotation around $1$-axis}.
\end{equation}
The algebra $A_W$ is generated by
\begin{equation}
 \varphi(f_W)\equiv \int d^4x~f_W(x) \varphi(x),
\end{equation}
or rather the unitaries $e^{i\varphi(f_W)}$, while $A_{W'}$ is generated by
\begin{equation}
 J_W e^{i\varphi(f_W)} J_W^{-1}= e^{-i\varphi(\mathcal{J}_Wf_W)},
\end{equation}
so that covariance is satisfied.

Since $[\varphi(x),\varphi(y)]=0$ if $x$ and $y$ are spacelike separated, causality is 
also fulfilled.

There is thus a consistent assignment of $A_W$ and $A_{W'}$: it is covariant and causal. 

Let us ignore the transverse coordinates $x^2$ and $x^3$ in test functions and study this 
localization further. Since, with $K_1\equiv K_W$,
\begin{equation}
\label{KW-operator-1}
 e^{it K_W}: ~ (x^0,x^1)~\to~(x^0\cosh t - x^1 \sinh t, -x^0 \sinh t + x^1 \cosh t),
\end{equation}
we have as $t \uparrow i\pi$,
\begin{equation}
 e^{-\pi K_W}:(x^0,x^1)\to (-x^0,-x^1).
\end{equation}

In quantum theory, $\mathcal{J}_W$ is represented by an anti-unitary operator $J_W$ and
\begin{equation}
 e^{-\pi K_W} \longrightarrow U(e^{-\pi K_W}) =  \Delta_W^{1/2}.
\end{equation}
Set
\begin{equation}
 S_W\equiv J_W \Delta_W^{1/2}.
\end{equation}

We remark that the continuation of $t$ to $i\pi$ requires a positive energy 
representation 
$U$. 
See \cite{Bal2016}.

The effect of $\mathcal{J}_W$ is compensated by $e^{- \pi K_W}$, so that $\mathcal{J}_W 
e^{-\pi K_W}$ acts as identity on $(x^0,x^1)$. Hence since $\varphi_W^*=\varphi_W$ 
($\varphi_W$ being a real field) and 
$\overline{f}_W=f_W$,
\begin{equation}
 S_W \varphi(f_W) S_W^{-1} = \varphi(f_W).
\end{equation}

We consider only free fields. Then since $\varphi(x)$ is linear in creation and 
annihilation operators, so is $\varphi(f_W)$ and
\begin{equation}
 \varphi(f_W)|0\rangle
\end{equation}
is a one-particle subspace.

Now, by (\ref{JW-operator}) and (\ref{KW-operator-1}),
\begin{equation}
 \mathcal{J}_W e^{it K_W} = e^{it K_W} \mathcal{J}_W,
\end{equation}
so that since $J_W$ is anti-unitary,
\begin{equation}
 J_W \Delta_W^{1/2} = \Delta_W^{-1/2} J_W
\end{equation}
and so
\begin{equation}
 S_W^2=\mathds{1}.
\end{equation}

Further, by the Lorentz invariance of the vacuum,
\begin{equation}
 J_W|0\rangle, ~~ \Delta_W^{1/2} |0\rangle, ~~ S_W |0\rangle \text{ are all }=|0\rangle.
\end{equation}

Thus if $\mathcal{H}$ is the one-particle Hilbert space of Fock space, 
$\varphi(f_W)|0\rangle$ is a ``real'' subspace $\text{Re} \mathcal{H}_W$ of 
$\mathcal{H}$:
\begin{equation}
 S_W \varphi(f_W)|0\rangle = \varphi(f_W)|0\rangle.
\end{equation}
It is real since $S_W$ being anti-linear, $i \varphi(f_W)|0\rangle$ does not belong to 
this subspace $\text{Re} \mathcal{H}_W$.

We can informally write
\begin{equation}
\label{Real-Part-of-Hil-in-W-1}
 \text{Re} \mathcal{H}_W = \frac{\mathds{1}+S_W}{2} \mathcal{H}.
\end{equation}

From $\text{Re} \mathcal{H}_W$ we can construct $A_W$ as Brunetti \emph{et al.} 
(cf. \cite{Bal2016}). 
discuss.

\vspace{0.3cm}

\emph{Summary}

\vspace{0.3cm}

 In the above we started by assuming that we have a free scalar field and arrived 
at $S_W$ and therefrom at $\text{Re} \mathcal{H}_W$. Since $\text{Re} \mathcal{H}_W$ also 
determines $A_W$, we now have an approach to localization where we start from the 
one-particle representation $\rho$ of the Poincar\'e group $\mathcal{P}_+$ on a complex 
Hilbert space $\mathcal{H}$. That supplies us with $S_W$ and hence $\text{Re} 
\mathcal{H}_W$ (\ref{Real-Part-of-Hil-in-W-1}). From this we recover $A_W$, the algebra of 
local observables in the wedge $W$.

This approach is more intrinsic as it starts just from Wigner's representation theory of 
the Poincar\'e group. It can also be applied to the case where the covariance group is 
the conformal group \cite{Longo}. It makes it clear that for localization in $W$ 
compatibly 
with Poincar\'e covariance and causality, we need the existence of $J_W$ and 
$\Delta_W^{1/2}=U\left(e^{-\pi K_W}\right)$.

\section{On the Infrared Effect}

We next consider a charged free massive scalar field $\varphi$ of charge $q$. In this 
case, the Fock space states get dressed by an infrared factor which breaks Lorentz 
invariance.

Let 
\begin{equation}
\label{charged-state-vector-photon-1}
 |0\rangle_\gamma |p\rangle
\end{equation}
denote the state vector when photon is in the ground state and the free charged particle 
has momentum $p$. When the interaction is switched on, 
(\ref{charged-state-vector-photon-1}) 
leads to an in state, namely
\begin{equation}
 |\text{in}\rangle \equiv \Omega ~ |0\rangle_\gamma |p\rangle,
\end{equation}
where the calculation of the dressing factor $\Omega$ is indicated below.

Since we are interested in very soft photons, we can ignore back reactions and treat the 
charged particle as moving with momentum $p$. Then the current of the charged particle is
\begin{align}
 J^\mu(x) &=  q\int d\tau~\delta^{(4)}(x-z(\tau))~\frac{dz^\mu}{d\tau}, \\
 z^\mu(\tau) &= \frac{p^\mu}{m}~\tau. \label{zmu-1}
\end{align}
The interaction term is thus
\begin{align}
  \int d^3x~A_\mu(x) J^\mu(x),
\end{align}
where $A_\mu$ is the electromagnetic potential. This leads to 
\begin{equation}
\label{dressing-factor-1}
 \Omega = \exp\left(-i q \int_{-\infty}^0 dx^0~\frac{p^\mu}{m}~A_\mu\left(\frac{p}{m}x^0 
\right) \right),
\end{equation}
upto factors unimportant for us. This $\Omega$ was worked out in \cite{Bal2016-2}.

We will work in the radiation gauge $A_0=\partial_i A^i = 0$ and in the interaction 
representation. Using the mode expansion of 
$A_i$,
\begin{align}
 A_i(x) &= \int d\mu({\bf k})~\left[a_i({\bf k}) e^{-ik\cdot x} + a_i^\dagger({\bf k}) 
e^{ik\cdot x}  
\right], \\
d\mu({\bf k})&= \frac{d^3 {\bf k}}{(2\pi)^{3/2} 2 k_0}, \\
\left[a_i({\bf k}),a_j^\dagger({\bf k}')\right] &= (2\pi)^{3/2}~ 2 k_0 
\left(\delta_{ij}-\hat{k}_i\hat{k}_j \right) \delta^{3}({\bf k}-{\bf k}'),
\end{align}
(with the rest of the commutators vanishing), we find
\begin{align}
 \Omega &=\exp\left(q \int d\mu({\bf k}) \left( a_i({\bf k})~\hat{\omega}^i(k)^+ - 
a_i^\dagger({\bf k})~ 
\hat{\omega}^i(k)^- \right) \right), \\
\hat{\omega}^i(k)^\pm &= \lim_{\varepsilon\downarrow 0} \frac{p_i-{\bf p}\cdot 
\hat{\bf k}~ \hat{k}_i  }{ k \cdot  p +i\varepsilon}.
\end{align}
But since $k\cdot p>0$ ($k$ is light-like with $k^0>0$ and $p$ is time-like with 
$p^0>0$), 
the $i\varepsilon$ can be dropped and we find dropping $\pm$ on $\hat{\omega}^i(k)^\pm$, 
that
\begin{equation}
 \Omega =\exp\left(q \int d\mu({\bf k}) \left( a_i({\bf k}) - a_i^\dagger({\bf k}) 
\right) \hat{\omega}^i(k) \right).
\end{equation}
Now, 
\begin{equation}
 \partial_0 A_i(x) = -i\int d\mu({\bf k})~ k_0~ \left[  a_i({\bf k}) e^{-ik\cdot x}- 
a_i^\dagger({\bf k}) 
e^{ik\cdot x}\right] = \text{Electric field } E_i.
\end{equation}
We will return to this equation a little later.

\vspace{0.3cm}

\emph{Interpretation of (\ref{dressing-factor-1})}

\vspace{0.3cm}

Equation (\ref{dressing-factor-1}) is the exponential of the Dirac-Wilson line integral, 
but in the time-like direction. Thus,
\begin{align}
 \Omega &= \exp\left(-i q \int_{-\infty}^0 dz^\mu A_\mu(z) \right),
\end{align}
where $z^\mu$ is given in (\ref{zmu-1}) with $\tau=x^0$, the time coordinate.

Under the gauge transformation
\begin{align}
 A_\mu ~ &\mapsto ~ A_\mu + \partial_\mu \Lambda,  \\
 \Omega ~ &\mapsto ~\Omega ~ e^{-i q \Lambda(0)} ~ e^{+i q \Lambda(-\infty)}.
\end{align}
This shows that $\Omega$ is created by a charge $q$ starting at time $-\infty$ and 
propagating to the origin at time $0$.

\section{The Boost in the Infrared Sector}

Let $K_i$ be the Lorentz boosts in the Fock space. For the electromagnetic field, they are
\begin{align}
 K_i &= \frac{1}{2} \int d^3x~x_i \left[{\bf E}(x)^2 + {\bf B}(x)^2 \right], \\
 B_i &=\varepsilon_{ijk} F_{jk},
\end{align}
where $F_{jk}=\partial_j A_k-\partial_k A_j$ and $E_i$ is the electric field conjugate to 
$A_i$:
\begin{equation}
 \left[A_i({\bf x},t), E_j({\bf y},t) \right]=i~\delta_{ij}^{T}({\bf x} 
-{\bf y}),
\end{equation}
where $\delta^T$ is the transverse $\delta$-function,
\begin{equation}
 \delta_{ij}^{T}({\bf x} -{\bf y}) = \left(\delta_{ij} - 
\frac{\partial_{x_i}\partial_{x_j}}{\partial_{\bf x}^2} \right) \delta^3({\bf 
x}-{\bf y}).
\end{equation}
Then,
\begin{equation}
\label{dressed-transformation}
 \Omega K_i \Omega^\dagger 
\end{equation}
acts on the in state vector. 

The electric and magnetic fields $E_i$ and $B_i$ are 
shifted by 
the transformation (\ref{dressed-transformation}). The shift of $E_i$ is
\begin{align}
 \delta E_i &= \left[q\int d\mu({\bf k}') \left(a_j({\bf k}') - a_j({\bf k}')^\dagger 
\right) 
\hat{\omega}^j(k'),~ i\int d\mu({\bf k}) k \left(a_i({\bf k})^\dagger e^{ik\cdot x} - 
a_i({\bf k}') 
e^{-ik\cdot 
x}\right) \right] \nonumber \\
     &= iq\int d\mu({\bf k})~ k~ \hat{\omega}_i(k) \left( e^{ik\cdot x} - e^{-ik\cdot 
x}\right) 
\nonumber \\
     &\equiv \omega_i(x)-\omega_i(-x). \label{E-variation}
\end{align}

A simple scaling argument shows that
\begin{align}
 \omega_i(\lambda x) &\underset{\lambda\to\infty}{\sim} \mathcal{O}\left( 
\frac{1}{\lambda^2}\right), \\
  & \text{or} \nonumber \\
 \omega_i(x) &\underset{x\to\infty}{\sim} \mathcal{O}\left( \frac{1}{x^2}\right).
\end{align}
Since
\begin{equation}
 \frac{1}{p\cdot k} = \frac{1}{p_0k_0 - {\bf p}\cdot {\bf k}}
\end{equation}
is not even in ${\bf k}$, we do not expect the $\mathcal{O}(1/x^2)$ term to cancel 
(\ref{E-variation}). With that 
assumption, we find the following term in $K_i$ to diverge logarithmically:
\begin{equation}
  \label{delta-E-square-1}
 \int d^3x ~ x_i ~ \delta \vec{E}(x)^2.
\end{equation}
After a cut-off, this term is positive.

If
\begin{equation}
 \Omega B_i \Omega^{-1} = B_i +\delta B_i,
\end{equation}
there is a similar contribution
\begin{equation}
 \frac{1}{2}\int d^3x x_i \delta \vec{B}(x)^2
\end{equation}
from the $\vec{B}^2$-term. As it is also non-negative, it cannot cancel 
(\ref{delta-E-square-1}).


In \cite{Balachandran2015}, the  divergence of $K_i$  is shown for vectors obtained by 
replacing omega by another  (``vertex'') operator . Also that paper focuses on the 
breakdown of Lorentz invariance and not localization.

There is a physical interpretation of the above result. A  Lorentz boost $\Lambda$ 
transforms  the photonic cloud of momentum $p$  into the photonic cloud with momentum 
$\Lambda p$.
A consequence of this transformation law is that  states of the coherent photon cloud do 
not belong to the domain of the infinitesimal generators of Lorentz boosts ${\bf K}$. The 
divergence found in the above calculation is also a proof of that behavior. An 
alternative argument can be obtained as follows. The expectation value of ${\bf K}$ in 
the photon cloud in particle mechanics is given by the sum of the contributions of each 
individual photon of the 
cloud. But that sum has the same degree of infrared divergence as 
\begin{equation}
 \langle N \rangle = \int d\mu({\bf k})~n_{\bf k},
\end{equation}
$n_{\bf k}$ being the number of photons in the cloud with momentum ${\bf k}$. This is in 
agreement 
with the previous result. Notice that on the contrary the same coherent  
quantum state  of the photon cloud belongs 
to the
domain of the QED Hamiltonian. Indeed,  once we renormalize the vacuum energy, the 
remaining energy is just the sum of the individual energies of each photon of the cloud
\begin{equation}
\mathcal{E}=\int d\mu({\bf k}) k_0 n_{\bf k} < \infty ,
\end{equation}
which is finite.

This concludes our argument that modular localization fails for charged fields.

There is of course a general argument \cite{Frolich1979} that Lorentz invariance breaks 
down for charged sectors 
of QED. That is enough to affirm the failure of standard localization arguments for 
charged particles. The merit of this paper is perhaps the fact that it is explicit.

\section{Remarks}

It has been argued elsewhere \cite{Bal2016-2} that non-abelian gauge theories, including 
QCD, 
breaks 
Lorentz invariance in sectors transforming non-trivially by the gauge group. Accordingly, 
standard localization arguments also fail in these sectors.

There is a striking resemblance between the Unruh effect and the physics of black holes. 
So we expect that our comments in this paper, which argue for the failure of 
localization arguments under generic conditions, have a bearing on the black hole 
information paradox.

\section{Acknowledgements}

A.P.B. and A.R.Q. thank S. Vaidya for discussions and for reminding us of refs. 1 and 2. 
The work of M.A. is partially supported by Spanish MINECO/FEDER grant 
FPA2015-65745-P and DGA-FSE grant 2015-E24/2. The work of A.R.Q. is 
supported by CNPq under process number 307124/2016-9.

\end{document}